# Grand Challenges for Embedded Security Research in a Connected World

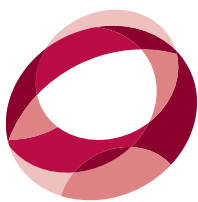

CCC
Computing Community Consortium
Catalyst

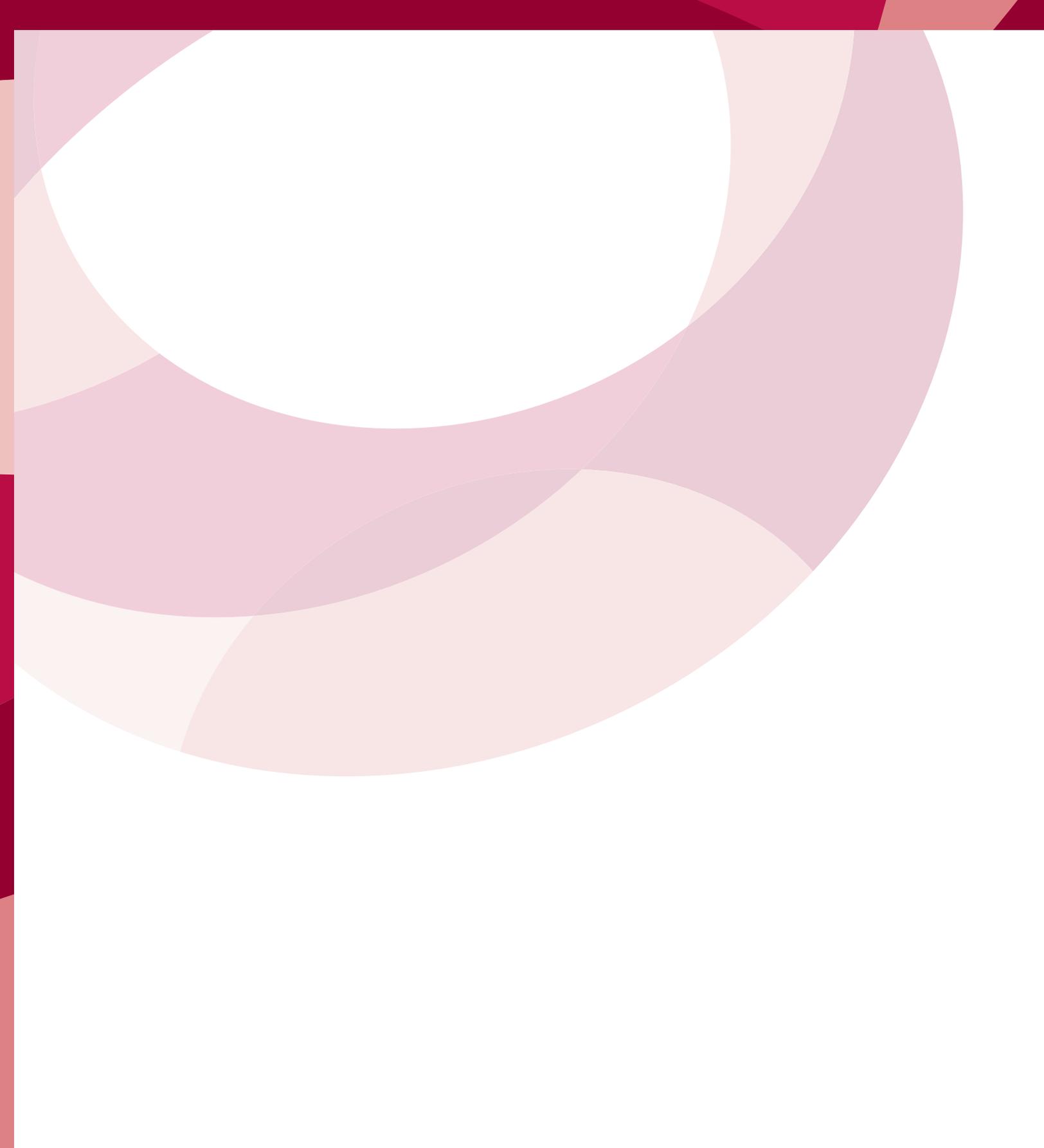
This material is based upon work supported by the National Science Foundation under Grant No. 1734706. Any opinions, findings, and conclusions or recommendations expressed in this material are those of the authors and do not necessarily reflect the views of the National Science Foundation.

# Grand Challenges for Embedded Security Research in a Connected World

A Report Based on a CCC Visioning Workshop Held August 12-13, 2018 in Baltimore, Maryland

**Authors:**

Denise Anthony
Wayne Burleson
Kevin Fu
Jorge Guajardo
Carl Gunter
Kyle Ingols
Jean-Baptiste Jeannin
Farinaz Koushanafar
Carl Landwehr
Susan Squires

Sponsored by the Computing Community Consortium (CCC)

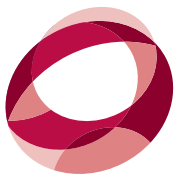

CCC
**Computing Community Consortium**
Catalyst

## 1. Executive Summary

Protecting embedded security is becoming an increasingly challenging research problem for embedded systems due to a number of emerging trends in hardware, software, networks, and applications. Without fundamental advances in, and an understanding of embedded security it will be difficult for future engineers to provide assurance for the Internet of Things (IoT) and Operational Technology (OT) in wide ranging applications, from home automation and autonomous transportation to medical devices and factory floors. Common to such applications are cyberphysical risks and consequences stemming from a lack of embedded security. The Computing Community Consortium (CCC) held a one-day visioning workshop to explore these issues. The workshop focused on five major application areas of embedded systems, namely (1) medical/wearable devices, (2) autonomous systems (drones, vehicles, robots), (3) smart homes, (4) industry and supply chain, and (5) critical infrastructure. This report synthesizes the results of that workshop and develops a list of strategic goals for research and education over the next 5-10 years. The full list of workshop participants can be found in the appendix.

Embedded security in connected devices presents challenges that require a broad look at the overall systems design, including human and societal dimensions as well as technical. Particular issues related to embedded security are a subset of the overall security of the application areas, which must also balance other design criteria such as cost, power, reliability, usability and function. Recent trends are converging to make the security of embedded systems an increasingly important and difficult objective, requiring new trans-disciplinary approaches to solve problems on a 5-10 year horizon.

Below is an overview of the major application areas and a brief summary of the research recommendations for each.

### Embedded Security Research Recommendations Across Major Application Areas

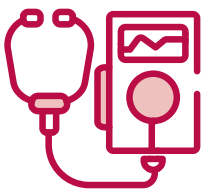

#### MEDICAL AND HEALTH DEVICES

Medical and health devices, both implanted and wearable are strictly regulated by the FDA for safety and effectiveness to balance the benefits to patient health against the risks from using any medical device. However, existing regulations for the safety and privacy (i.e., HIPAA) issues related to medical devices do not cover information security or cyberphysical attack situations. Furthermore, fitness and personal health monitoring devices present numerous vulnerabilities and are not currently regulated.

**Challenges:**
- Long legacy tail makes it challenging to change or update system interfaces or add new procedures (such as authentication protocols).
- Severe power and energy constraints of wearable, mobile, and implantable medical devices.
- Software, when seen as a medical device, intersects and sometimes conflicts with the existing regulatory structures such as HIPAA.
- Globalization and distribution of medical devices away from the countries of origin.

**Research Recommendations:**
The application of classic cryptography, security, and control theory (which can be used to model and study impact of various attacks on cyber-physical systems as well to mitigate damage) to the vulnerabilities and attack surfaces could yield novel solutions. Stronger authentication protocols for devices that leverage unique features related to the physics, locality, or possibly distance to/from device. Ongoing efforts to create devices and systems with appropriate failback solutions and safe-modes can enable innovative applications while also limiting the risks.

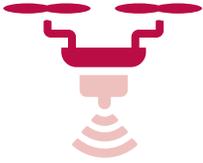

## DRONES AND TRANSPORTATION

Our traditional modes of transportation, such as cars and airliners, are increasingly computerized, connected, and thus vulnerable to cyber-attacks. At the same time, these modes of transportation are more and more autonomous, from cars to public transportation to (potentially) flying taxis. Autonomy also enables the emergence of new, smaller logistical capabilities such as flying drones for package delivery.

**Challenges:**
- These transportation systems directly interact with the physical world, in many cases have real-time requirements, and have the capability to harm people.
- Transportation systems have long lifespans on the order of decades with multiple patching and testing cycles. Note that some (eg. Tesla) are now pushing software updates however this is far from widespread. Software updates can also introduce vulnerabilities.
- Sensors in these systems are easier to spoof than human eyes.
- How do you share and manage electronic keys and consumer data?

**Research Recommendations:**
Develop a methodology and tools (including formal methods) that incorporate security from the conception of the vehicle and enable reasoning about multiple layers (control, software, hardware) with different assumptions. This methodology should also be able to leverage interactions among multiple layers or physical properties of existing systems to enhance the security of the overall vehicle.

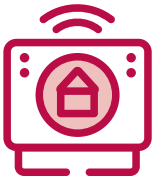

## SMART HOMES

Currently, embedded home systems operate technologies from simple light switches that can be turned on with a cell phone, to integrated home fire alarms, security alerts, and health monitoring systems (such as sensors that detect falls). These home systems operate in conjunction with third parties including smart electric power meters from the electricity service provider, or mobile systems that are found on private automobiles, which may share controls with the garage door opener. Such embedded technologies are less regulated and more likely to be operated by a non-professional.

**Challenges:**
- Operation of the system by non-professionals with little knowledge of security requires a robust system that does not rely on outside intervention for configuration.
- Current technologies are not always integrated though there is a growing emphasis on standardization. For instance, the fire alarm and door lock systems may be from different vendors and fail to communicate with one another. Additionally, the increasing number of home devices, managed in different ways and with different user interfaces, is more likely to increase confusion among users, and hence introduce security flaws that are exploitable.
- As home embedded systems become more capable they create new types of flows of information. These include not only the data collected from, say, a security system with cameras, but also devices with voice controls that listen at all times for keywords that trigger their functions.
- The richness of the system is also likely to create fresh types of side channels such as the ability to use fluctuations on the power system to detect, say, the program being watched on the home TV.
- The information collected from individuals in their homes has obvious privacy implications and users will expect a high level of security to protect their sensitive information.

**Research Recommendations:**
It is necessary to develop some kind of *rely-guarantee framework* (a kind of concurrent software verification technique) in which components can announce their security properties along with the assumptions they expect from their environment. Answers to the following questions must be generated through regulations: Who should patch or update devices to assure continued protection? Should the lifetime of devices match those of home ownership? Or, should there be models of transfer of devices to the new owner?

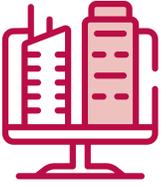

### INDUSTRY AND SUPPLY-CHAIN

Embedded systems rely heavily on software and firmware, but even more so they rely on the hardware that executes the code and makes the system real. Due to the long lifetime of industrial and supply-chain systems, some of this hardware is so old that new parts are nearly impossible to come by and replacement parts must be purchased from third parties, with varying degrees of success and fidelity. Even when new systems are built, they are often beholden to legacy interfaces for the sake of interoperability.

**Challenges:**
- Old systems and protocols are challenging to secure retroactively.
- Old hardware is similarly challenging to secure, and also challenging to acquire securely. Due to their age, they are often highly resource constrained, leaving little headroom to accommodate updated software or firmware with modern cryptographic and defensive technologies.
- Designing a safe and secure modern Application-Specific Integrated Circuit (ASIC) is challenging and expensive, often costing up to 100 million dollars[1.]

**Research Recommendations:**
Retaining the capability to manufacture new parts is a key solution to the threat of counterfeits. Ideally, a vendor that is no longer interested in manufacturing a part — or worse, a vendor that goes out of business altogether — should be required to yield their design to others who may wish to do so. For newer designs, so-called "split ASIC" and multi-chiplet techniques can divide the design into separate pieces that can be sent to separate fabrication facilities, complicating an adversary's efforts.

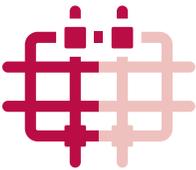

### SMART GRID AND CRITICAL INFRASTRUCTURE

Electric meters once passively recorded the accumulated flow of current into a household or business and were read monthly. Smart meters now record and report power consumption on a second-by-second (or finer) basis in real time, permitting charging based on time of use. More recently, "smart grid" has broadened to incorporate grids in which the infrastructure includes a range of technologies that can more generally sense and control its own operation. These new abilities can be used to take advantage of distributed power generation based on renewable sources (solar panels, wind turbines), to provide earlier detection and location of outages, and to control power flow among regions more safely and efficiently.

**Challenges:**
- Traditional centralized power generation is becoming more distributed as smaller scale generation with decentralized ownership and control (often "behind the meter") becomes more economical. More active management of power demand is likely to accompany this transition, but unexpected dependencies between infrastructures are likely to be revealed, particularly in emergency situations. Failures and attacks may propagate in unexpected ways and pricing will become more dynamic.

**Research Recommendations:**
Educating power companies about the effects of their buying decisions is crucial, and appropriate application of cryptographic technologies can solve some problems in this domain. Cryptography can assure the integrity of control signals even if they pass through untrusted domains, for example.

---

[1] DARPA, "Circuit Realization at Faster Timescales (CRAFT)" *https://www.darpa.mil/program/circuit-realization-at-faster-timescales*

## 2. Introduction

This report is based on a 2018 Computing Community Consortium (CCC) visioning workshop and describes results from the workshop including conclusion and reflections afterward by a subset of the participants. The report describes the workshop objectives, format, and participants, and then report the results of five breakout groups in each of five application areas. These results are then summarized with common themes and differences discussed, concluding with research recommendations for the next 5-10 years.

The Leadership in Embedded Security Workshop was sponsored by the CCC and co-located with the USENIX Security Symposium in Baltimore, Maryland on August 13, 2018. Kevin Fu (University of Michigan), Wayne Burleson (University of Massachusetts Amherst), and Farinaz Koushanfar (University of California San Diego) organized and chaired the workshop with the help of CCC staff, Khari Douglas and Ann Schwartz Drobnis.

Approximately 50 experts in embedded security were invited following their response to a broad and general call for participation. Participants were selected based on a 1-page position paper that addressed a grand challenge in embedded security with a 5-10 year horizon across multiple subdisciplines. Participants came from academia (including foreign academic participants from the United Kingdom, Switzerland, Belgium, China and Korea), industry, and government agencies within the United States.

The workshop program consisted of introductions, two keynotes, two panels, and afternoon breakout groups. The first keynote, by Sam Fuller, CTO emeritus of Analog Devices, discussed the history of embedded security and emphasized the importance of simplifying and minimizing the trusted computing base. The second keynote, by Farinaz Koushanfar of UC San Diego, discussed recent work in machine learning and its role in embedded security. Following the keynotes, there were two panel presentations with Q&A sessions. The first panel featured international perspectives on research support for academics from four countries (the United Kingdom, the Swiss Confederation, the Republic of Korea, and the People's Republic of China), while the second panel offered perspectives from United States government agencies (FDA, DHS, DoD, NSF) on the roles and priorities for embedded security research within the U.S.

**The afternoon program broke up the workshop participants into five groups to discuss embedded security in the following major application areas:**

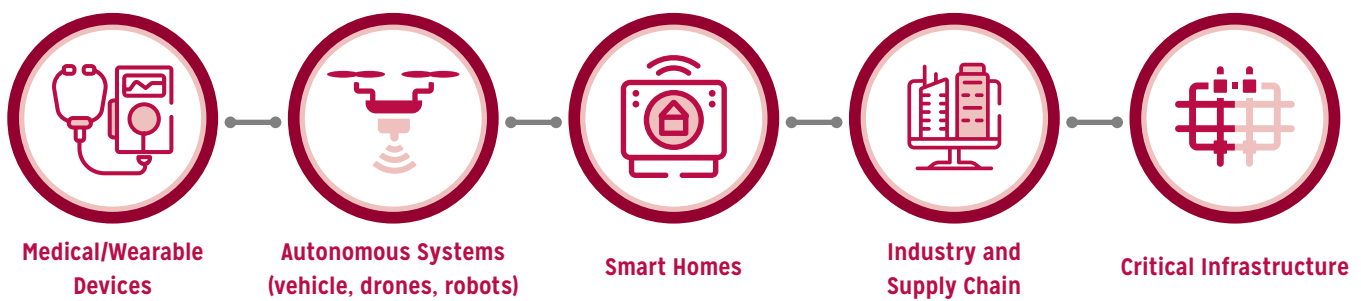

**Medical/Wearable Devices** — **Autonomous Systems (vehicle, drones, robots)** — **Smart Homes** — **Industry and Supply Chain** — **Critical Infrastructure**

Each group was given the charge of identifying key trends, challenges, and solutions in their respective areas. At the end of the session, each group reported back on their discussions, and their findings are presented in the next section of the report.

# 3. Embedded Security by Major Application Areas
## 3.1 Medical and Wearable Devices

Medical devices are strictly regulated by the FDA for safety and effectiveness to balance the benefits to patient health against the risks that comes from using any medical device. However, existing regulations for safety and privacy issues (i.e., HIPAA) related to medical devices do not cover information security or cyber attack situations. The newly proposed FDA Medical Device Safety Action Plan[2] includes a recognition that the exploitation of device security vulnerabilities could threaten their safety or effectiveness, and thus the cybersecurity of devices must be considered and managed. **A key challenge going forward is balancing the interconnected but distinct issues of safety, cybersecurity, and usability of medical devices.** The **long legacy tail** of many devices and clinical systems is a particular challenge for cybersecurity because there are limited options to change or update system interfaces, or to add new procedures (such as authentication protocols). This also creates a burden to maintain the training and expertise necessary to manage these devices and systems. Another **challenge for cybersecurity of medical devices is the sometimes severe power and energy constraints of wearable, mobile and implantable medical devices.** This is due to a combination of battery size and battery life as well as thermal constraints, especially for implantable devices. Recent research has explored various methods of energy harvesting (chemical, mechanical, thermal).

A number of current trends intersect with the particular challenges of ensuring the (cyber)security of medical devices. The continuing growth of **software as a service in the cloud** must consider the implications for *software as a medical device*, which is defined by the International Medical Device Regulators Forum (IMDRF) as "software intended to be used for one or more medical purposes that perform these purposes without being part of a hardware medical device."[3] Software as a medical device **intersects and sometimes conflicts with the existing regulatory structures such as HIPAA.** The already high levels of variability in patient/user populations (e.g., elderly, children, soldiers) will increase as more medical devices are introduced to locations outside of traditional medical environments (e.g., homes, battlefields), increasing not only the **complexity of usability challenges, but also the threats to security**. Furthermore, it is likely that over time more patients will have multiple devices that work in parallel but all intersect with embedded infrastructure and systems. A related issue is the increasing development and spread of wellness applications and devices that are not regulated by existing regulatory structures (e.g., FDA, HIPAA) like certain smartwatches and fitness trackers.

As the supply chain for components of medical devices connected to embedded systems becomes increasingly globalized the number of security challenges also increase. Two additional global trends with implications for medical devices are: (1) medical tourism, in which patients have a medical procedure or get a device outside of the United States (U.S.) but must maintain it over time in the U.S., and/or patients who receive treatment in the U.S. but live elsewhere; and (2) shipping of (sometimes used) devices to other countries that have fewer resources to maintain/update different system capabilities and standards.

Ongoing research and development in a number of key areas is necessary to address these trends and challenges, and to ensure the future security of medical devices. Given the relatively recent emergence of embedded and connected medical devices, the application of classic cryptography, security, and control theory (which can be used to model and study impact of various attacks on cyber-physical systems as well to mitigate damage) to the vulnerabilities and attack surface could yield novel solutions. Another promising direction for development is toward new, stronger authentication protocols for devices that leverage unique features related to the physics, locality, or possibly distance to/from device. Finally, continuing ongoing efforts to create devices and systems with appropriate failback solutions and safe-modes can enable innovative applications while also limiting the risks.

---

[2] FDA, "Medical Device Safety Action Plan: Protecting Patients, Promoting Public Health". 2019.
 *https://www.fda.gov/aboutfda/centersoffices/officeofmedicalproductsandtobacco/cdrh/cdrhreports/ucm604500.htm*

[3] FDA, "Software as a Medical Device (SaMD)" 2018 *https://www.fda.gov/MedicalDevices/DigitalHealth/SoftwareasaMedicalDevice/default.htm*

## 3.2 Transportation: from planes, trains and vehicles, to drones and beyond

Our traditional modes of transportation, such as cars and airliners, are increasingly computerized and connected, and thus vulnerable to cyber-attacks. At the same time, these modes of transportation are more and more autonomous. Autonomy has enabled the emergence of, smaller logistical capabilities such as flying drones for package delivery. Manufacturers are becoming aware of the risks posed by the possibility of cyberattacks and are adopting traditional IT security techniques to protect themselves. However, the nature and criticality of their security is distinct from IT security, and therefore require distinct solutions. These transportation systems directly interact with the physical world, in many cases have real-time requirements, **have the capability to harm numerous people beyond the user**, and have long lifespans on the order of decades with multiple patching and testing cycles.

When assessing the cybersecurity of transportation, it is **important to carefully characterize the risk.** For example, there are many ways to cause an accident and harm the occupants of one car without using cyber techniques. However, it is much more difficult to cause hundreds of accidents at the same time using only physical means. Sensors progressively replace human eyes, which is concerning from a security point of view as a sensor is easier to spoof. Another recent trend has been the consolidation of many components into a smaller number of processors (e.g. in the Tesla model of embedded computation), causing security concerns as it creates a **single point of failure for the vehicle.** Electronic keys are also a concern: how do you manage and share them while at the same time preventing thefts and attacks? Finally, vehicles, in particular consumer cars, generate a lot of data, and it is currently unclear who owns this data and how to address the **privacy of the data if it is owned by the manufacturer.**

To address all of these challenges, it is necessary to continue developing a methodology and tools (including formal methods) that incorporate security from the conception of the vehicle and transportation system, and enable reasoning about multiple layers (control, software, hardware) with different assumptions. One particularly pressing example might be the need to develop tools to verify the security of embedded software for transportation and find security vulnerabilities in an automatic or semi-automatic manner. Similarly, there is a trend towards integrating typical IT security solutions (firewalls, intrusion detection systems) into automotive domains. On the one hand, this leads to the detection of typical attacks seen also in the enterprise domain. On the other hand, **these techniques need to be adapted to cope with the real-time requirements of transportation systems.** This methodology should also be able to leverage interactions among multiple layers or physical properties of existing systems to enhance the security of the overall vehicle. Such an approach will benefit from multiple security checks and thus, not a single point of failure in the system design. It will also lead to the development of new platform architectures. **In addition, benchmarks and metrics** to evaluate security of different kinds of vehicles should be introduced. Although it is recognized that security is notoriously hard to measure, metrics can be developed based on restricted threat models and the economics of both attack and defense. Finally, it is necessary to develop security regulations and economic incentives for transportation, in the same spirit as existing safety regulations (eg. seat-belts, air-bags, tire-pressure sensors).

## 3.3 Smart Homes Technologies

As embedded smart home technologies expand in use and complexity, they will become more and more vulnerable to security risks. Currently, embedded home systems operate technologies that range from simple light switches that can be turned on with a cell phone, to integrated home fire alarms and security alerts, and interfaces to health monitoring systems such as sensors that detect falls. These home systems operate in conjunction with third party devices and applications including smart electric power meters from the electricity service provider, and the mobile systems that are found on private automobiles, which may share controls with the garage door opener. Such embedded technologies are less regulated and more likely to be operated by a non-professional.

**Four key security risks of home embedded systems have been identified.**

1. Operation of the system by non-professionals with little knowledge of security requires a robust system that does not rely on outside intervention for configuration.

2. Current technologies are not always integrated, though there is a growing emphasis on standardization. For instance, the fire alarm and door lock systems may be from different vendors and fail to communicate with one another. Additionally, the increasing number of home devices, managed in different ways and with different user interfaces, is more likely to increase confusion among users, and hence introduce security flaws that are exploitable.

3. As home embedded systems become more capable they create new types of flows of information. These include not only the data collected from, say, a security system with cameras, but also devices with voice controls that listen at all times for keywords that trigger their functions. Recent smart speakers and voice controlled devices have allowed users great convenience but at the expense of reduced privacy and perhaps security.

4. Opportunities to use malware and exploits are already emerging. We have seen the use of home embedded systems for denial of service (DoS) attacks in the Mirai botnets. It is only a matter of time before more sinister attacks such as ransomware appear in this domain. The richness of the system is also likely to create fresh types of side channels such as the ability to use fluctuations on the power system to detect, say, the program being watched on the home TV. These go beyond protecting direct access to the data sources which may have access control and/or encryption.

Several trends create, drive, or shape these key challenges. Non-professional users will require increased convenience and ease of use from these products. Unlike traditional WIMP (window / icon / mouse / pointer) user interfaces, the next generation of embedded home systems will feature voice- and gesture-based command and control. These new user interfaces create new risks for the injection of commands into the system - for example sounds made by other devices in the home or external, or concealed commands through popular songs on smart speakers (e.g. the Burger King "Ok Google" commercial or inaudible voice commands like DolphinAttack)[4,5].

Responding to user needs, one trend is toward product integration among vendors. This includes high-level scripting languages like If This Then That (IFTTT) and the development of hubs like Samsung SmartThings, which is able to support an app store and common device programming interfaces. However, these systems may present new vulnerabilities. Finally, we need to consider the growing numbers of people who will participate in embedded home systems. For instance, in the past a homeowner may have entrusted a neighbor with the spare house key. Now the homeowner may offer access rights to a monitoring service, that same neighbor, their children and so on through smart devices. In particular, cell phone and cloud services take home systems to the general Internet rather than just operating locally, thus offering a large attack surface and a very wide potential attacker community.

One key point is that the security of individual devices does not assure the security of the combination of these devices in all home configurations. Ultimately it is necessary to develop some kind of rely-guarantee framework in which components can announce their security properties along with the assumptions they expect from their environment. For instance, many devices will expect security from physical tampering to be provided by another system like a door lock. Other systems, like the electric power meter attached to the outside of the house, will have some level of physical tamper resistance on their own. Embedded systems for homes could also benefit from Privacy Enhancing Technologies (PETs) that collect selected, but not all, data about home devices, such as electric use that might lead to ideas for reducing costs or coordination with others to address emergencies. The PETs would need to be similarly protected from tampering and illegitimate access.

---

[4] Jacob Kastrenakes, "Burger King's new ad forces Google Home to advertise the Whopper," The Verge. 2017. *https://www.theverge.com/2017/4/12/15259400/burger-king-google-home-ad-wikipedia*

[5] Guoming Zhang et al., DolphinAttack: Inaudible Voice Commands. 2017. *https://arxiv.org/pdf/1708.09537.pdf*

There are other questions that we anticipate with the lifecycle of home embedded devices. For example, **who should patch or update devices to assure continued protection?** Should the lifetime of devices match those of home ownership? Or, should there be a model of transfer of devices to the new owner? Resolving such issues should enhance security and privacy, reduce expenses, and offer welcome convenience in what might be a chaotic process of ownership transfer.

## 3.4 Industrial and Supply Chain

Embedded systems rely heavily on software and firmware, but even more so they rely on the hardware that executes the code and makes the system real. Due to the long lifetime of these systems, some of this hardware is so old that new parts are nearly impossible to come by and replacement parts, with varying degrees of success and fidelity, must be purchased from third parties. Even when new systems are built, they are often beholden to legacy interfaces for the sake of interoperability.

**Old systems and protocols are challenging to secure retroactively.** For example, it is common for industrial supervisory control and data acquisition (SCADA) systems to use the almost 40 year old ModBus protocol for communications. Standards of that era are often built for *safety*, not *security*, and are not easily updated to adapt to modern threats and threat models.

Old hardware is similarly challenging to secure, and also challenging to acquire securely. Due to their age, they are often **highly resource constrained, leaving little headroom to accommodate updated software or firmware with modern cryptographic and defensive technologies.** Often the manufacturers no longer offer the parts, forcing users to go to third parties to acquire replacements in case of failure. As the U.S. Government Accountability Office (GAO) found, it is challenging to acquire genuine parts in this manner[6]. An adversary selling counterfeit parts for profit poses a safety risk, but an adversary selling counterfeit parts for malice may pose a security risk as well.

Replacing the hardware isn't easy, either. Modern process nodes (e.g. 7nm FinFETs) are only available from a trio of vendors, which poses the normal risks of a limited marketplace. In addition, small feature sizes often motivate large, complicated designs with billions of transistors. **Designing a safe and secure modern ASIC is challenging and expensive,** often costing up to 100 million dollars[7].

Retaining the capability to manufacture new parts is a key solution to the threat of counterfeits. **Ideally, a vendor that is no longer interested in manufacturing a part — or worse, a vendor that goes out of business altogether — should be required to yield their design to others who may wish to do so.** This "design escrow" concept should permit safe new production of otherwise obsolete parts, perhaps on process nodes that remain behind the state of the art. Being behind the curve, yet ahead of the original part, provides a "sweet spot" that mitigates the threats of the 7nm troika while still providing some benefits of smaller feature sizes vs. the original design.

---

[6] GAO, "DOD Suply Chain: Suspect Counterfeit Electronic Parts Can Be Found on Internet Purchasing Platforms" 2012. *https://www.gao.gov/products/GAO-12-375*

[7] DARPA, "Circuit Realization at Faster Timescales (CRAFT)" *https://www.darpa.mil/program/circuit-realization-at-faster-timescales*

For newer designs, a variety of approaches can be employed to mitigate the threat. So-called "split ASIC" and multi-chiplet techniques can divide the design into separate pieces that can be sent to separate fabrication facilities, complicating an adversary's efforts. Where performance concerns permit, designs should aggressively employ microcode and firmware to permit post-fabrication tuning of the hardware. Designs should also leave sufficient "unused headroom" to accommodate new software features, both security and otherwise. Such systems are easier to patch over time, and may even be able to use firmware to work around flaws discovered later in the hardware. Finally, business incentives as well as standards can force suppliers to provide upgrade paths.

### 3.5 Smart Grid and Critical Infrastructure

For many people, the term "smart grid" refers to an electric power infrastructure in which there are "smart meters." Electric meters once passively recorded the accumulated flow of current into a household or business and were read monthly; smart meters record and report power consumption on a second-by-second (or finer) basis in real time, permitting billing based on time of use. They may also exert some control over power flow. The fine-grained measurement of power consumption raises significant privacy issues and has been studied in great detail by Kevin Fu et al.[8]

The added sensing and control can enable better modeling of demand, and thus improve the efficiency and safety of the overall power system and make it easier to incorporate renewables, such as solar panels and wind turbines, which are inherently variable. Added sensing and control can also be used to reduce peak power demands and thus reduce the need for expensive and inefficient "peaker plants" to accommodate peak demand. Demand smoothing can be encouraged either via pricing, for example, so that consumers have an incentive to operate a dishwasher late at night when other demands are low, or enforced via actual remote control of "smart" appliances.

Traditional centralized power generation is becoming more distributed as smaller scale generation with decentralized ownership and control, often "behind the meter," becomes more economical. More active management of power demand is likely to accompany this transition, and pricing will become more dynamic. Unexpected dependencies between infrastructures are likely to be revealed, particularly in emergency situations, and failures and attacks may propagate in unexpected ways. Bitcoin miners seeking the cheapest power have already imposed significant demands in unexpected locations, triggering pricing reviews. Adoption and deployment of smart grid technology may also vary globally in surprising ways. Countries with established and redundant infrastructures will need to accommodate substantial legacy systems, while less developed nations may be able to adopt smart grid technologies more quickly.

Grid systems traditionally were segregated from communication systems. Telephone networks were self-powered so they would continue to operate when the power grid went down. Consumers were strictly segregated from the generation and distribution systems. Operated as utilities, these systems invested in high reliability components with long lifetimes, much as the telephone systems of the AT&T telephone monopoly days did. When the telephone monopoly was ended, new companies began producing components to be integrated into the existing telephone networks, and switching system internal interfaces suddenly became much more widely accessible. Customers buying rather than leasing phones looked for low initial cost, features, and style more than reliability and security. These preferences continue to influence many aspects of the evolution of the telephone ecosystem.

However, in addition to large utilities, the US power grid also features rural cooperatives, with different tradeoffs of budgets and engineering. Indeed, these coops are where one might first see use of Internet and commodity tablet computers and smartphones to control the grid.

---

[8] Andrés Molina-Markham et al,, Private memoirs of a smart meter. 2010. *https://dl.acm.org/citation.cfm?doid=1878431.1878446*

Consumers will gain abilities to generate power and control consumption; "smart" components with shorter life expectancy will be incorporated into networks with aging but reliable "dumb" components. Sensing and signaling system interfaces will be opened up. A major challenge will be to maintain power system dependability and security as these changes percolate through these systems. As the set of users becomes wider, authentication and authorization, particularly in emergency situations, will be critical. Cyber attacks will become more feasible, particularly if low cost, unpatchable, or only manually patchable systems permeate the grid.

Indeed, this might be the crucial security issue with the smart grid: the vastly increased attack surface. Educating power companies about the effects of their buying decisions is crucial. A way that consumers (including power companies) may make informed decisions about the likely dependability and security effects of their purchases is very much needed. One important aspect of realizing this goal is to ensure that the costs of a security or dependability failure are borne by the companies in a position to address that failure. In this case, the costs of failures will be internalized and developers will have the appropriate incentives to bring dependable and secure products to market. Past and present policies for software production and licensing are not good models in this regard. Anderson raised this issue as early as 2001, however it is still largely true[9].

Appropriate application of cryptographic technologies can solve some problems in this domain. Cryptography can assure the integrity of control signals even if they pass through untrusted domains, for example. As always, careful attention will need to be focused on the key management and protocols involved. Other critical issues include balancing high security with low latency on critical (and often slow legacy) communication paths, and balancing the security shelf-life of key lengths and algorithms with the decades-long deployment lifetimes of grid equipment. Industry standardization and rigorous analysis, testing, and certification of components should be encouraged.

## 4. Common and Distinguishing Themes and Solutions

The workshop breakout presentations revealed that there were numerous common themes in terms of trends, challenges, and solutions that impacted most or all of the five applications areas. However, there were also some important distinctions between the areas. For example, increased connectivity leads to increased attack surfaces in all five areas, however network-based defenses vary considerably depending on the threat models and impacts on performance and utility. As another example, machine learning algorithms and the move towards autonomous systems are particularly significant in all areas, especially for detecting intrusion and anomalous behavior, however specific vulnerabilities vary depending on the area. Furthermore, the requirements (performance, cost, reliability, life-time, etc.) for each area are significantly different, thus leading to different challenges, and warranting different approaches to security and privacy. Figure 1 shows the overlap of common trends and challenges by application area as well as notable areas of distinction, while Figure 2 shows areas of overlapping, potential novel solutions for these challenges.

---

[9] For more on this topic see chapters 22 and 25 in Ross Anderson's *Security Engineering: A Guide to Building Dependable Distributed Systems*.

| Common Themes and Challenges to Embedded Security | Smart Homes | Medical | Transportation Systems & Autonomous Vehicles | Industrial Control | Supply Chain | Smart Grid & Infrastructure |
|---|---|---|---|---|---|---|
| Long Life Time | ✓ | ✓ | ✓ | ✓ | ✓ | ✓ |
| Need to Accommodate Legacy Systems | ✓ | ✓ | ✓ | ✓ | ✓ | ✓ |
| Classic Control Theory | ✓ | ✓ | ✓ | ✓ | ✓ | ✓ |
| Consumer Privacy | ✓ | ✓ | ✓ |  |  | ✓ |
| Increasing Autonomy |  |  | ✓ | ✓ |  |  |
| Power & Energy Constraints |  | ✓ |  |  |  | ✓ |
| Authentication in Emergency Situations | ✓ | ✓ | ✓ | ✓ |  | ✓ |
| Globalization | ✓ | ✓ | ✓ |  | ✓ |  |
| Diverse Users & User Skill Level | ✓ | ✓ | ✓ |  |  |  |
| Software as the Service in the Cloud | ✓ | ✓ |  |  |  |  |
| Cloud Increased Multi-Vendor Interoperation | ✓ | ✓ | ✓ |  |  | ✓ |
| Concentrated Intellectual Property |  |  |  | ✓ | ✓ |  |

*Figure 1. Overlap of Common Themes and Challenges in Embedded Security by Application Area*

| Potential Novel Solutions to Embedded Security Challenges | Smart Homes | Medical | Transportation Systems & Autonomous Vehicles | Industrial Control | Supply Chain | Smart Grid & Infrastructure |
|---|---|---|---|---|---|---|
| Appropriate fallback solutions & safe modes | ✓ | ✓ | ✓ | ✓ | ✓ | ✓ |
| Leveraging physics, locality, and distance for authentication | ✓ | ✓ | ✓ |  |  |  |
| Application of classic control theory to control those systems |  |  | ✓ | ✓ | ✓ | ✓ |
| Formal methods to incorporate security from the beginning | ✓ | ✓ | ✓ | ✓ | ✓ | ✓ |
| Developing benchmarks & metrics to evaluate security & safety | ✓ | ✓ | ✓ | ✓ | ✓ | ✓ |
| Security regulations & economic incentives | ✓ | ✓ | ✓ | ✓ | ✓ | ✓ |
| Design a privacy dashboard for users to understand their data | ✓ | ✓ | ✓ |  |  |  |
| Model & secure the full lifecycle of IoT devices | ✓ | ✓ | ✓ | ✓ | ✓ | ✓ |
| Split design - leverage slower trusted parts to mediate technology executed in less trusted parts |  |  |  | ✓ | ✓ |  |

*Figure 2. Overlap of Potential Novel Solutions to the Embedded Security Challenges of Each Application Area*

## 4.1 Common Themes

Several common IT trends, both technological and societal, cut across all areas. For example, embedded systems are leveraging the ever advancing capabilities of semiconductor technology, incorporating numerous processors, memories, radios and sensors in low-cost and low-power sub-systems. In addition, there is a move in all industries towards the integration of multiple and diverse functionalities into single devices to reduce the complexity of systems and their associated costs. While this is not widespread, it is driven by start-ups (rather than well-established players), which typically aim to disrupt entire ecosystems. Thus, in the next five to ten years we can expect to have heterogeneity in deployed solutions, with conceptually different designs. At the same time, these different solutions and design paradigms will need to co-exist and guarantee common sets of security and privacy assurances, individually targeted to each industry. Embedded systems also reflect typical globalization trends, with design, manufacturing, and deployment occurring across geographic and political boundaries. Systems are increasingly designed with large numbers of outsourced components, both hardware and software. These lead to concerns regarding the security of the components and how they will influence the security of the systems, many critical infrastructure, in which they will be integrated.

Threat models are also changing with new capabilities and business models for cyber criminals. Threat actors may range from nation states to petty criminals, and even include the users themselves (e.g. un-locking features or using unauthorized code). Users of embedded systems now encompass larger and more diverse portions of the population, with wide economic, education and cultural differences. Autonomy is a general trend in most embedded systems, as users delegate more and more functionality to embedded processors often assisted by AI and machine learning techniques. It is expected that this trend towards depending more on AI algorithms will only intensify in the future. This in turn will lead to AI algorithms being increasingly responsible for (safety) decisions that will impact the well-being of humans and that were traditionally taken by humans. There is a need to define what the "correct" decisions are in many life-or-death scenarios and, more generally, to define algorithms that can learn such decision making protocols, which until now where exclusively the responsibility of humans. Balancing tradeoffs between security and safety, cost, usability, and rapid design development are also common across the five areas.

The role of the life-cycle of systems is another recurring theme in embedded security. This ranges from design and manufacturing to field updates, key management, cryptographic agility, data retention, changes in ownership, and changes in threat models. System lifetime is a critical aspect shared by all five of the major application areas, ranging from the design and manufacture of systems, through their deployment, numerous upgrades, and finally decommissioning and disposal of systems and their associated data. Life-cycle management becomes even more relevant when we observe that these widely deployed embedded systems will collect and have access to highly confidential (business) data, privacy sensitive data, or key material that can grant access to critical infrastructure or safety critical systems.

Most embedded systems need well designed fail-safe mechanisms and resiliency in case of attack. For medical and automotive systems, an attack can quickly lead to fatalities, so manual overrides are necessary. However these need to be carefully designed to avoid new vulnerabilities. Critical infrastructure, by its nature, requires strong resiliency plans, although the time-frame for response is longer than that for medical systems. The sheer scale of critical infrastructure makes this resiliency a difficult challenge in system design, however solutions can be shared with those for other catastrophic scenarios (weather, earthquake, terrorism, etc.).

## 4.2 Distinguishing Themes

Several important distinctions were noted between the five application areas. A primary point of distinction is in the ownership of the system and the overall economic situation. For example, a smart home owner wants the ability to control the security and privacy of their personal life, while an autonomous vehicle owner may have to comply with more regulations due to the shared nature of roadways. The owner of a power distribution network has an ongoing business relationship with their customers, as well as significant regulations for both safety and fairness.

The owner of the data that the system generates also varies considerably for a medical device compared to critical infrastructure. The device manufacturer may claim ownership of some data as a way to improve the device performance. Modern data-driven businesses will want as much data as possible to optimize their systems for users and profit, as well as to protect against various threats. Ultimately, users should have rights to the data generated from their bodies and their behavior. Laws such as the recent European Union General Data Protection Regulation (GDPR), and market incentives can be used to help ensure these rights. In contrast, the data from critical infrastructure will be collected by local, state, and federal governments, as well as manufacturers and installers in order to monitor usage and design improvements to systems. Data from critical infrastructure will need to be protected from misuse but must be made available to those stakeholders who manage them.

The role of regulations and incentives differs considerably across the five areas as well. Even within the area of medical and wearable devices, there is a huge range between FDA-regulated devices and the almost unregulated world of personal health apps on mobile phones. The data on both types of devices can be highly personal and thus a significant privacy risk. The FDA, FTC or some other government agency should consider developing regulations for these new consumer devices and associated software.

Critical infrastructure such as power, transportation, or communications typically involves a relationship between a large provider and numerous clients. The provider has substantial resources to afford complex security solutions, however the economics of a large user base may make it difficult to protect client security and data privacy. Regulations or market incentives may be necessary to reconcile this tension.

Incentives can be more effective than regulations in some embedded security solutions. As security and privacy become more visible and valuable to users, hopefully business incentives can play more of a role in setting security policy.

Threat models also vary between application areas. Industry supply chains are threatened by counterfeits and recycled/repurposed parts that result in lost revenue and damage to brand reputation. Critical infrastructure threats include users who are not paying for services delivered and third-party analytics firms who misuse client data. The more obvious threats of terrorism and nation-state warfare loom large in critical infrastructure where an attack can have widespread impact and denial of service. In these cases, resilience and recovery are probably more realistic goals given the difficulty of preventing attacks.

**Finally, it can be useful to recognize that embedded security problems can be classified across several axes. These include the following:**

### ⬤ AXIS 1

**Area of Impact:** Personal vs. Community

**Point of Tension:** The security and rights of the individual or client need to be balanced with the overall security of the system. For example, an individual could refuse to update their operating system, but this may introduce vulnerabilities into the system if the lack of updates also prevents them from receiving the same quality of security upgrades.

### ⬤ AXIS 2

**Area of Impact:** Products vs. Services

**Point of Tension:** This refers to the embedded systems as well as their security. Lately the economy has seen a shift from products (mostly static systems, e.g. a car) to services (ongoing, non-tangible benefits available for purchase, e.g. updating navigation system in a car), which both complicates attacks as well as allowing updates and threat-sharing.

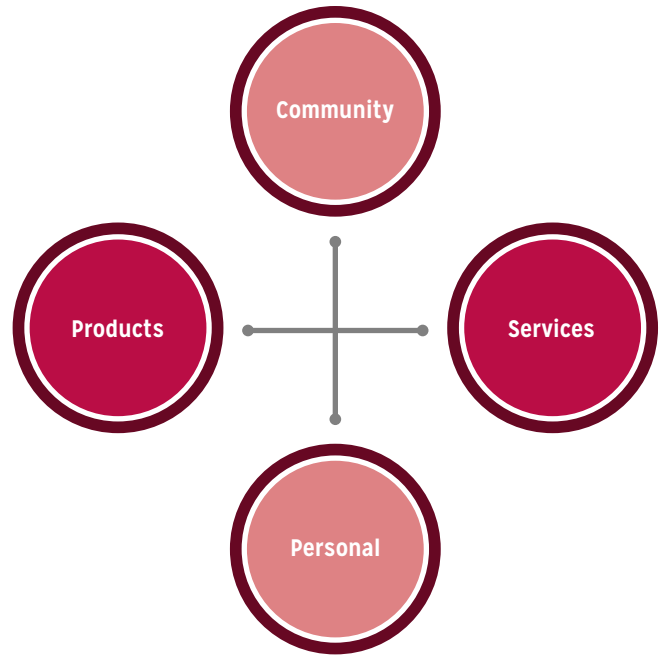

### ⬤ AXIS 1

**Area of Impact:** Consumer vs. Producer/Vendor

**Point of Tension:** The security of the consumer in terms of protecting their data versus the concerns of the manufacturers of security solutions and their concerns to protect themselves in terms of liability, and also their intellectual property and security of their own systems. For instance, what should individuals be required to share about the status of their smart homes? This data could reveal lots of personal information, but could also be necessary for the provider to find and fix vulnerabilities.

### ⬤ AXIS 2

**Area of Impact:** Regulations vs Incentives

**Point of Tension:** Both can be useful in terms of setting and enforcing security policies. As threat models change, they need to be revisited and adapted – some will require more of the carrot (incentives) and some more of the stick (regulations).

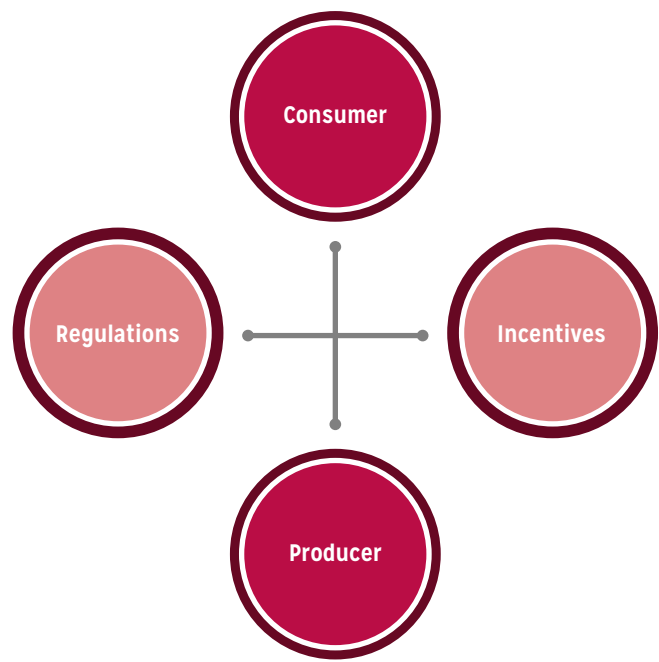

## 5. Directions and Growth areas in Embedded Security Research and Education

Potential solutions to the embedded security challenges discussed above can be divided across three action areas as follows.

### 5.1 Cryptographic Schemes and New Security Methods

These include new cryptographic techniques among which three should be highlighted. First, quantum computers are expected to become practical in the next 10-20 years and this will require the design and validation of new so-called post-quantum cryptographic schemes. Such schemes are based on mathematical problems for which no efficient quantum algorithms are known to exist. As such, they are expected to remain secure for appropriate parameter sizes even if quantum computers become practical in the medium term. Widely varying new algorithms are based on lattices, multi-variate methods, hashes, codes and elliptic curve isogeny. NIST is already developing a first standard for Post-Quantum Cryptography[10] with drafts expected in 2022-2024, but this will be an ongoing effort since most of these schemes are quite young and will presumably evolve. In January 2019, the CCC held a workshop on Post Quantum Cryptography – see that workshop report for further information on this topic.

Second, the advent and wide deployment of machine learning algorithms coupled with the broad collection of data and ever increasing security concerns and vulnerabilities have prompted renewed interest in techniques that allow processing of data in the encrypted domain (i.e., without having to decrypt the data). Two different lines of work should be highlighted here. On the one hand, there has been intense research since 2009 on Fully Homomorphic Encryption (FHE) Schemes; these are cryptographic schemes with homomorphic properties. In other words, computation can be performed directly on encrypted data without decrypting. Here the focus and current challenges are on making such schemes practical. Practicality first entails the ability to perform the operations with acceptable overhead and reduce the size of key material, which tend to be significantly larger than standard (not post-quantum secure) schemes such as RSA or elliptic curves.

While FHE is applicable to protect remote computation on sensitive data, multi-party computation (MPC) is a more efficient scheme for computation in a group of mutual distrusting parties. These schemes are computationally less costly per device but they require the exchange of significant amounts of encrypted data among the parties involved in the computation. Recently the most promising solutions are based on a combination of the previously mentioned techniques, using FHE and MPC to perform operations that are best suited for each. It is also important to point out that the basic ideas underlying MPC have found applicability in other domains such as secure supply chain and production of chips and the design of cryptographic schemes at the hardware level that offer security against physical (tampering) adversaries.

In addition to these cryptographic techniques, the security of existing and future solutions can only be assured by a combination of formal methods aimed at providing strong security guarantees about deployed systems, and the detailed consideration of human factors in their design. Particularly important are formal methods to verify software and hardware designs as well as techniques aimed at semi-autonomously finding bugs in software and hardware designs, in a wide range of user scenarios. Given the expected increase in our societal dependence on autonomous systems the relevance of such techniques will only increase. Different types of users and their preferences should be studied when designing secure systems and during research, leveraging a wide body of human factors research in related areas.

Another area of interest is the design and implementation of cryptographic schemes leading to small footprints and acceptable performance in highly constrained devices. The increased levels of connectivity and data gathering will be enabled by the expected deployment of sensor and actuator nodes with very limited computational capabilities. Thus, designing for such systems continues to be an interesting challenge.

---

[10] NIST, "Post-Quantum Cryptography" 2020. *https://csrc.nist.gov/projects/post-quantum-cryptography*

Finally, the research community has started to explore the use of economic models of security and economic incentives for private entitituess to improve security outcomes (e.g. taking down fradulent web-sites), often working in partnership with other stakeholders such as law enforcement. An entire sub-discipline of Information Security Economics[11] has emerged in the last decade to explore these cross-disciplinary issues,

## 5.2 Practical Security Solutions

Security solutions evolve continuously and so existing solutions to new challenges are being deployed across industries. Examples include the design of secure software upgrades for cyber physical systems, the automotive domain being a prime example of this type of solution and associated challenges. Similarly, there has been an increase in the number of tools offered that help developers write secure software by integrating them into their development tool chains. These tools are made easy to use for all developers, not only security experts. In addition, there is a trend towards the development of tools that will detect and patch vulnerabilities automatically without, or with very little, human intervention.

## 5.3 Education Initiatives

Finally, we emphasize the need for educating the workforce in the basics of secure design and how to apply standard guidelines to the building of secure products. Similarly, it is necessary to educate developers on the use of privacy by design techniques to guarantee that both security and privacy safety measures are built into products from their onset and not simply as an afterthought. Building without privacy for design is destined for failure, as has been shown so often in the past (recently with Facebook, dating app Coffee Meets Bagel, and the city of Tallahassee).[12]

Users of products would also benefit from general security and privacy education campaigns aimed at making them aware of simple secure configuration options for home devices, typical scams performed via email or social media, the perils of sharing too much information, and practicing simple digital hygiene, all of which can go a long way towards protecting their personal information. Such approaches are already being actively deployed in companies with the aim to minimize security incidents. It is clear that the general population would benefit from such an approach as well.

---

[11] The Workshop on the Economics of Information Security *https://econinfosec.org/*

[12] Steve Turner, "2019 Data Breaches - The Biggest Breaches of the Year," Identity Force. 2019. *https://www.identityforce.com/blog/2019-data-breaches*

# 6. Appendix

## 6.1 Summary of Recommendations:

Below is a list of high-level areas for potential research, education, and policies that could eliminate or reduce some of the risks discussed in this workshop report. These concepts are indexed by topic area. This report aims to capture the discussions that occurred at the August 2018 workshop—because of this certain topics are not discussed in the report, but are still listed below as areas to consider for improving embedded security.

**Cross-Application Areas Research Areas:**

◗ **Practical**
- Design methodologies – page 1, 9, 12, 17-18, 22-23
- Forensics
- Secure upgrades – page 4, 11, 22-23
- User interfaces – page 1, 4, 7, 10, 11, 14

◗ **Scientific**
- Economics models – page 9, 19
- Formal methods – page 2, 9, 23
- Lightweight and robust implementations – page 3, 9
- New cryptography (PQC, FHE, MPC) – page 1, 2, 8, 14, 22
- User-studies – page 23

**Domain-specific Research Areas:**

◗ **Autonomous Systems – page 2, 8, 15, 18, 23**
- Complex composite systems and interoperability with legacy
- Threat models – page 8
- User vs. manufacturer incentives – page 18

◗ **Industrial and Supply-chains – page 4, 11-12, 15, 19**
- Economics of countermeasures vs. threats of counterfeiting – page 4, 12, 19

◗ **Medical – page 1, 6-8, 9, 15, 18**
- Human safety and privacy foundations – page 6, 8, 9, 18

◗ **Smart Grid and Critical Infrastructure – page 4, 12-14, 15-16, 16-19**
- Data-driven studies of large-scale system effects with user inputs

◗ **Smart Homes – page 3, 9-11, 15, 21**
- Life-cycle issues – page 11
- User-driven decisions about risk tolerance – page 21

**Education Areas:**

◗ **Enforcement – page 23**
- Analysis – page 14
- Containment
- Detection – page 9
- Forensics

◗ **Policy makers**
- Adaptive policies (based on results)
- Regulations and incentives – page 2-3, 6, 9, 18-22

◗ **Users – page 10, 11, 14, 17, 18, 19, 24**
- Cost – page 19
- Digital hygiene – page 24
- Security and privacy awareness – page 24
- Tradeoffs – page 19

◗ **Workforce – page 24**
- Beyond Computer Science and IT – anyone who will design, analyze, regulate or enforce security and privacy in embedded systems
- Pipeline

**Common solutions:**

- Application of classic control theory to control those systems – page 15
- Automated tools (eg software analysis) – page 2, 9, 24
- Crypto agility – page 22-23
- Design escrow – page 12
- Economic/business incentives – page 9, 13, 14, 18-23
- Emerging cryptography (MPC, FHE, PQC,…) – page 22-23
- Key management – page 14, 17, 23
- Leveraging the physics, locality, and distance for authentication – page 2, 8
- Lightweight HW security primitives
- Open source security libraries, software, and hardware
- Secure updates, mixed-lifetime components, field-replaceable unit – page 3, 11, 13-14, 15-20
- Resiliency, attack recovery, containment – page 18-19
- Split design (not just semiconductors) – page 4, 12
- User education – page 17, 24
- Workforce education – page 24

## 6.2 Workshop Participants

| Name | Affiliation |
|---|---|
| Ross Anderson | University of Cambridge |
| Denise Anthony | Dartmouth & University of Michigan |
| Reza Azarderakhsh | Florida Atlantic University and PQSecure Technologies |
| Ryan Burchfield | National Security Agency |
| Wayne Burleson | University of Massachusetts Amherst |
| George Burrus | University of South Florida |
| Kevin Bush | MIT Lincoln Laboratory |
| Srdjan Capkun | ETH Zürich, Switzerland |
| Todd Carpenter | Adventium Lab |
| Charles Clancy | Virginia Tech |
| Sauvik Das | Georgia Tech |
| Robert Dick | University of Michigan |
| Khari Douglas | Computing Community Consortium |
| Ann Drobnis | Computing Community Consortium |
| Michael Dunaway | NIMSAT Institute University of Louisiana at Lafayette |
| Brian Fitzgerald | Food and Drug Administration |
| Kevin Fu | University of Michigan |
| Sam Fuller | CTO Emeritus of Analog Devices |
| Daniel Genkin | University of Pennsylvania |
| Jorge Guajardo | Robert Bosch LLC – Research and Technology Center |
| Carl Gunter | University of Illinois |
| Peter Harsha | Computing Research Association |
| Dan Holcomb | University of Massachusetts Amherst |
| Ken Hoyme | Boston Scientific |
| Kyle Ingols | MIT Lincoln Laboratory |
| Jean-Baptiste Jeannin | University of Michigan |
| Benjamin Justus | Siemens Corporation |
| Yongdae Kim | Korea Advanced Institute of Science and Technology |
| Farinaz Koushanafar | University of California, San Diego |
| Sandip Kundu | National Science Foundation |
| Carl Landwehr | George Washington University |
| Insup Lee | University of Pennsylvania |
| David Maimon | University of Maryland |
| Dan Massey | University of Colorado Boulder |
| Douglas Maughn | Department of Homeland Security |
| Howard Meyer | Department of Defense |
| Miroslav Pajic | Duke University |
| Brad Reaves | North Carolina State University |
| Mastooreh Salajegheh | Visa Research |

| Name | Affiliation |
|---|---|
| Hassan Salmani | Howard University |
| Armin Sarabi | University of Michigan |
| Patrick Schaumont | Virginia Tech |
| Sean Smith | Dartmouth College |
| Susan Squires | University of North Texas |
| Edward Suh | Cornell University |
| Tomas Vagoun | NITRD |
| Ingrid Verbauwhede | KU Leuven – COSIC & UCLA |
| Dongyan Xu | Purdue University |
| Wenyuan Xu | Zhejiang University |
| Yuval Yarom | University of Adelaide |

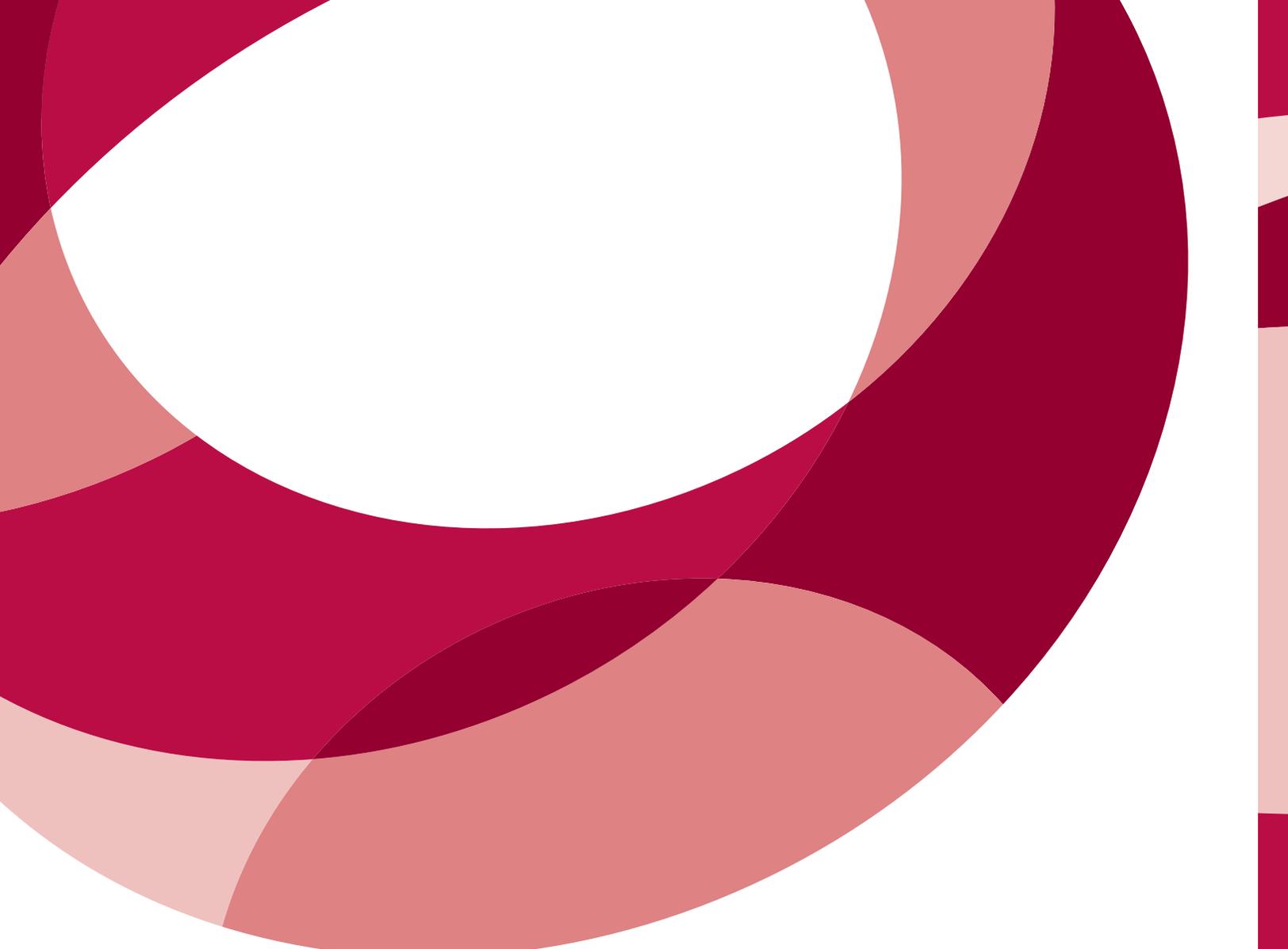
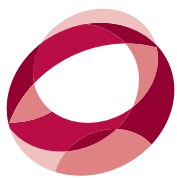
CCC
Computing Community Consortium
Catalyst

1828 L Street, NW, Suite 800
Washington, DC 20036
P: 202 234 2111 F: 202 667 1066
www.cra.org cccinfo@cra.org